\documentclass{ws-ijmpb}

\usepackage{savesym}
\usepackage{latexsym,amsmath,amsfonts,amssymb}
\savesymbol{captionbox}
\usepackage{caption}
\restoresymbol{CPT}{captionbox}

\def\vev#1{\left\langle #1 \right\rangle}

\begin{document}

\markboth{Carlos Hoyos}
{Hall viscosity, topological states and effective theories}

%
%

\title{HALL VISCOSITY, TOPOLOGICAL STATES AND EFFECTIVE THEORIES}


\author{CARLOS HOYOS}


\address{Raymond and Beverly Sackler Faculty of Exact Sciences \\
School of Physics and Astronomy \\
Tel-Aviv University, Ramat-Aviv 69978, Israel.\footnote{choyos@post.tau.ac.il}}




\maketitle


\begin{abstract}
Hall viscosity is a dissipationless transport coefficient whose value is quantized in units of the density in some topological phases and may be used as a measure of topological order. I give an overview of the Hall viscosity, its relation to Hall conductivity in Galilean invariant theories and its realization in effective theories.
\end{abstract}


\section{Introduction}

The Hall viscosity,\cite{Landau10,Avron1995} similarly to the Hall conductivity, is a dissipationless transport coefficient that is allowed in two spatial dimensions when time reversal invariance and parity are broken. While the Hall conductivity appears in the current, the Hall viscosity is a term in the stress tensor. 

The interest in the Hall viscosity is motivated by its topological nature. In certain gapped systems such as Quantum Hall fluids and $p$-wave superfluids the value of the Hall viscosity over the density is determined by the `orbital angular momentum per particle' or the shift, the change in magnetic flux when the system is put in a sphere.\cite{Read2009,Read2011} Deformations that do not close the gap and that preserve rotational invariance do not affect to the value of this ratio. Furthermore, a measurement of the Hall viscosity can distinguish among different states with the same filling fraction, which is measured using the Hall conductivity. In particular, the Hall viscosity can be non-zero in charge neutral states where the Hall conductivity vanishes, so it serves as an alternative characterization of topological order.

Even though it would be clearly interesting to measure the Hall viscosity, there are no experimental results as yet. A difficulty compared to the Hall conductivity is that time-dependent stresses are clearly more difficult to measure than static currents. There is however hope to overcome this issue, in principle it is also possible to measure the Hall viscosity using inhomogeneous electric fields \cite{Hoyos2012,Bradlyn2012} or X-ray diffraction by phonons.\cite{Phonon}

Besides these considerations, an interesting question in theoretical physics is which systems have a quantized Hall viscosity over density ratio, and how it is determined. The relation to the orbital spin that was computed microscopically is confirmed in effective theories of non-relativistic systems. Relativistic systems on the other hand have proven to be more elusive, there is no known example where the Hall viscosity is non-zero at zero temperature, but for some gauge/gravity models. A closely related term has been found in the canonical energy-momentum tensor in the presence of torsion (a non-symmetric affine connection).\cite{Hughes2011,Hughes2013} This term does not appear in the symmetric energy-momentum tensor, but can be relevant in the effective description of solids with dislocations and broken time reversal invariance. The questions about the Hall viscosity remain as an interesting field that is still open to exploration.

In this note I will give an overview of the subject, including some recent developments. I have tried to make a balanced presentation and be pedagogical. This has forced me to omit many details in order to keep the length of the text within reasonable bounds. There enters some personal bias, in general there is more space dedicated to continuum effective theories and symmetries and less to microscopic derivations, specially adiabatic calculations, which are nevertheless very important. I hope the references provided will be sufficient to satisfy the interested reader. There are probably other unintended omissions for which I apologize in advance. In order to simplify notation I work with units $\hbar=c=e=k_B=1$.

I will start in \S \ref{sec:def} by reviewing the definition of Hall viscosity. In \S \ref{sec:top} I will explain the relation of the Hall viscosity to topological properties in Hall fluids and chiral superfluids. In \S \ref{sec:kub} I will present Kubo formulas for the Hall viscosity and its relation to Hall conductivity. In \S \ref{sec:hyd} I discuss the Hall viscosity in effective theories.   I summarize the main points in \S \ref{sec:con}.

\section{A non-dissipative viscosity} \label{sec:def}

The main properties of the Hall viscosity and its physical effects in fluids can be found in the early review Ref.~\refcite{Avron1997}.  For completeness, we partly review them here as well. 

A small deformation parametrized by a displacement vector $\xi_i$, $i=1,\dots,d$  produces a stress that depends on the strain $\xi_{ij}=\partial_i \xi_j+\partial_j \xi_i$ and the strain rate $\dot{\xi}_{ij}\equiv \partial_t\xi_{ij}$ through the elastic modulus ($\lambda$) and viscosity ($\eta$) tensors
\begin{equation}
T_{ij}=p\delta_{ij}-\lambda_{ijkl}\xi_{kl}-\eta_{ijkl}\dot{\xi}_{kl}.
\end{equation}
The first term is the pressure. For a fluid the only response to a strain is due to changes of volume
\begin{equation}
\lambda_{ijkl}=\kappa^{-1}\delta_{ij}\delta_{kl},
\end{equation}
where $\kappa^{-1}$ is the inverse compressibility that determines changes in the pressure with volume
\begin{equation}
\kappa^{-1}=-V \frac{\partial p}{\partial V}.
\end{equation}
We will mostly deal with fluids. When time reversal invariance is not broken, the viscosity tensor satisfies Onsager's relations
\begin{equation}\label{onsager}
\eta_{ijkl}=\eta_{klij},
\end{equation}
which for a rotationally invariant system allows only two possible transport coefficients, the shear ($\eta$) and bulk ($\zeta$) viscosities
\begin{equation}
\eta_{ijkl}=\eta(\delta_{ik}\delta_{jl}+\delta_{il}\delta_{jk})+\left(\zeta-\frac{2}{d}\eta\right)\delta^{ij} \delta_{kl}.
\end{equation}
When time reversal invariance is broken, as for instance if a background magnetic field is turned on, the conditions \eqref{onsager} are relaxed and it is possible to have an `odd' contribution to the viscosity\cite{Landau10,Avron1995}
\begin{equation}\label{antisyodd}
\eta^{(A)}_{ijkl}=-\eta^{(A)}_{klij}.
\end{equation}
A peculiarity of the odd viscosity is that is dissipationless. The variation of the energy density under a strain is
\begin{equation}
\delta \varepsilon = -T_{ij}\delta \xi_{ij}.
\end{equation}
Using the first law of thermodynamics $\delta \varepsilon = T\delta s-p\delta V$, with $s$ the entropy density, $T$ the temperature and $V$ the volume, the change of entropy with time becomes
\begin{equation}
T\dot{s} = \eta_{ijkl} \dot{\xi}_{ij}\dot{\xi}_{kl}.
\end{equation}
Due to \eqref{antisyodd}, the contribution proportional to the odd viscosity vanishes. This implies that the odd viscosity can be non-zero even at zero temperature.

In general, $\eta^{(A)}=0$ if rotational invariance is not broken. However, for $d=2$ spatial dimensions an odd viscosity is allowed if parity is also broken
\begin{equation}\label{hallvisc}
\eta^{(A)}_{ijkl}=-\frac{\eta_H}{2}(\epsilon^{ik}\delta^{jl}+\epsilon^{jk}\delta^{il} +\epsilon^{il}\delta^{jk}+\epsilon^{jl}\delta^{ik}).
\end{equation}
We label this as Hall viscosity (although it may appear in systems without magnetic fields). An alternative way to see that the Hall viscosity is non-dissipative is to use a no-force condition. At finite density the value of the Hall conductivity can be derived in this way. For a system with a current $J^\mu$ coupled to the electromagnetic field $F_{\mu\nu}$, the conservation equation of the energy-momentum tensor is
\begin{equation}
\partial_\alpha T^{\alpha\mu}=F^{\mu\alpha}J_\alpha.
\end{equation}
The no-force conditions are
\begin{equation}
E_i J^i=0, \ \ \bar{n} E_i = B\epsilon_{ij}J^j.
\end{equation}
Where we have written the expressions in terms of the magnetic field $F_{ij}=B\epsilon_{ij}$ and the density $J^t=\bar{n}$. The first condition is satisfied if $J^i=\sigma_H \epsilon^{ji} E_j$. Plugging this in the second condition one finds the classical value of the Hall conductivity
\begin{equation}
\sigma_H =\frac{\bar{n}}{B}.
\end{equation}
The Hall viscosity can be derived from a similar analysis, by coupling the system to an external metric. The conservation equation of the energy-momentum tensor changes to
\begin{equation}
\nabla_\alpha T^{\alpha\mu}=\partial_\alpha T^{\alpha\mu}+\Gamma^\alpha_{\alpha\beta} T^{\beta\mu}+\Gamma^\mu_{\alpha\beta} T^{\alpha\beta}=F^{\mu\alpha}J_\alpha,
\end{equation}
where $\Gamma^{\alpha}_{\mu\nu}$ are the Christoffel symbols. Let us do a small time-dependent geometric deformation which preserves the volume. The spatial metric is changed to
\begin{equation}
g_{ij}=\delta_{ij}+h_{ij}(t), \ \ \delta^{ij} h_{ij}=0.
\end{equation}
Then, the no-force condition for $E_i=0$ becomes
\begin{equation}
\Gamma^t_{ij}T^{ij}=0, \ \ T^{tk}\Gamma^i_{tk}-B\epsilon^{ik} J_k=0.
\end{equation}
In this case the current and the momentum can be set to zero $T^{ti}=J^i=0$. The general solution for the stress tensor is
\begin{equation}
T^{ij}=p\delta^{ij}-\frac{1}{2}\eta^{ijkl}\partial_t h_{ij}, \ \ \eta^{ijkl}=-\eta^{klij}.
\end{equation}
The tensor $\eta$ is the Hall viscosity. Note that in contrast to the Hall conductivity, its value is not fixed by the no-force condition.

\subsection{Relativistic Hall viscosity}

The Hall viscosity can be generalized to relativistic theories. Aside from its intrinsic interest, relativistic hydrodynamics in 2+1 dimensions can be used as an effective description of graphene.\cite{Mueller:2008,Fritz:2008,Mueller:2009} A complete discussion of relativistic hydrodynamics with parity-breaking terms can be found in Ref.~\refcite{Jensen:2011xb},\footnote{For a recent discussion of non-relativistic parity-breaking hydrodynamics see Ref.~\refcite{Kaminski2013}.} we follow here some of their notation. Let us define the three-velocity $u^\mu$, $\mu=0,1,2$ with components 
\begin{equation}
u^0 = \gamma, \ \ u^i= \gamma v^i,
\end{equation}
where $v^i=\dot{\xi}_i$, $i=1,2$ is the spatial velocity, and $\gamma=1/\sqrt{1-v^2}$. The three-velocity is normalized so that it has unit norm respect to Minkowski's metric with mostly plus convention ($\eta_{00}=-1, \; \eta_{ij}=\delta_{ij}$)
\begin{equation}
\eta_{\mu\nu} u^\mu u^\nu =-1.
\end{equation}
The energy-momentum tensor of a relativistic fluid takes the form
\begin{equation}
T^{\mu\nu}=(\varepsilon+p)u^\mu u^\nu + p \eta^{\mu\nu}+\pi^{\mu\nu},
\end{equation}
where $\varepsilon$ is the energy density fluid and $\pi^{\mu\nu}$ contains higher derivative terms. In the Landau frame, defined by the equation
\begin{equation}
T^{\mu\nu}u_\nu=-\varepsilon u^\mu,
\end{equation} 
the first order derivative terms preserving Lorentz invariance and parity are the relativistic shear and bulk viscosities
\begin{equation}
\pi^{\mu\nu}=-\eta \sigma^{\mu\nu} - \zeta P^{\mu\nu}\partial_\alpha u^\alpha,
\end{equation}
where
\begin{equation}
\sigma^{\mu\nu}=P^{\mu\alpha} P^{\nu\beta}\left(\partial_\alpha u_\beta+\partial_\beta u_\alpha-P_{\alpha\beta}\partial_\sigma u^\sigma\right),
\end{equation}
and $P^{\mu\nu}=\eta^{\mu\nu}+u^\mu u^\nu$ is the projector transverse to the velocity.

The Hall viscosity term takes the form
\begin{equation}
\pi^{\mu\nu}_{\rm odd} = \frac{\eta_H}{2} \left(\epsilon^{\mu\alpha\beta}u_\alpha\sigma_\beta^{\ \nu}+\epsilon^{\nu\alpha\beta}u_\alpha\sigma_\beta^{\ \mu}\right).
\end{equation}
In fact this is not the only possible term that can appear to first order in derivatives, but other contributions are scalar (as the bulk viscosity) rather than tensor terms.

As in the non-relativistic case, the Hall viscosity is dissipationless because it does not contribute to the divergence of the entropy current $J_s^\mu=s u^\mu$. One can show this easily. First, we project the conservation equation of the energy-momentum tensor along the three-velocity:
\begin{equation}
0=\partial_\mu T^{\mu\nu} u_\nu = -\partial_\mu \varepsilon u^\mu -(\varepsilon+p)\partial_\mu u^\mu +\frac{\eta_H}{2} \left(\epsilon^{\mu\alpha\beta}u_\alpha \partial_\mu \sigma_\beta^{\ \nu}u_\nu+\epsilon^{\nu\alpha\beta}\partial_\mu u_\alpha\sigma_\beta^{\ \mu} u_\nu\right)+\cdots.
\end{equation}
The dots refer to the shear and bulk viscosity contributions. We now apply the first law of thermodynamics,
\begin{equation}
\varepsilon+p=T s, \ \ d\varepsilon = T ds,
\end{equation}
and we obtain the following expression for the divergence of the entropy current
\begin{equation}
T \partial_\mu(s u^\mu)=\frac{\eta_H}{2} \left(\epsilon^{\mu\alpha\beta}u_\alpha \partial_\mu \sigma_\beta^{\ \nu}u_\nu+\epsilon^{\nu\alpha\beta}\partial_\mu u_\alpha\sigma_\beta^{\ \mu} u_\nu\right)+\cdots.
\end{equation}
A bit of algebra shows that
\begin{align}
&\epsilon^{\mu\alpha\beta}u_\alpha \partial_\mu \sigma_\beta^{\ \nu}u_\nu = -\epsilon^{\mu\alpha\beta}u_\alpha \partial_\mu u^\lambda \partial_\lambda u_\beta+\epsilon^{\mu\alpha\lambda} u_\alpha\partial_\mu u_\lambda \partial_\sigma u^\sigma,\\
&\epsilon^{\nu\alpha\beta}\partial_\mu u_\alpha\sigma_\beta^{\ \mu} u_\nu = -\epsilon^{\mu\alpha\beta}u_\alpha \partial_\beta u^\lambda \partial_\lambda u_\mu-\epsilon^{\mu\alpha\lambda} u_\alpha\partial_\mu u_\lambda \partial_\sigma u^\sigma.
\end{align}
Adding the two terms we find that the contribution of the Hall viscosity to the divergence of the entropy current vanishes.

\section{Hall viscosity in topological phases} \label{sec:top}

We have seen that the Hall viscosity is not dissipative, so it will generically be non-zero in (2+1)-dimensional systems where parity and time-reversal symmetry are broken, even at zero temperature. In the classification of weakly coupled systems,\cite{Kitev2009,Schnyder2009} there are three possible subclasses of topological systems\footnote{In principle only time reversal symmetry and particle-hole symmetry are used, a more refined classification taking into account parity can be found in Ref.~\refcite{Chiu2013}.} that we have grouped in table~\ref{table2}. 

\captionsetup{width=.75\textwidth}
\begin{table}[h]
\begin{center}
\begin{tabular}{|c|c|c|}
\hline Class & PHS & Spin \\  \hline
A   & 0 & - \\ \hline
C & +1  & singlet\\ \hline
D &  -1 & triplet\\ \hline
\end{tabular}
\caption{\small Classes of two-dimensional topological insulators/superconductors with broken time reversal invariance according to particle-hole symmetry (PHS). $0$ means no symmetry and $\pm 1$ are different realizations of the symmetry. For the classes $C$ and $D$, PHS distinguishes whether the Cooper pair is a spin singlet or triplet.} \label{table2}
\end{center}
\end{table}

\begin{itemize}
\item Type A: Quantum Hall (non-zero magnetic field $B$) and Chern insulator (zero magnetic field)\cite{Haldane1998} enter in this category.
\item Type C: $d$-wave superfluids/superconductors ($d_{x^2-y^2}+id_{xy}$)\cite{Senthil1998,Senthil1999,Gruzberg1999}.
\item Type D: Chiral superfluids/superconductors ($p_x\pm i p_y$)\cite{Read2000}. 
\end{itemize}
In interacting systems the Integer Quantum Hall phase generalizes to fractional phases and chiral $p$- and $d$-wave superfluids can in principle be generalized to anyon superfluids with fractional orbital angular momentum. 

The calculation of the Hall viscosity for free electrons in a magnetic field with completely filled Landau levels was carried out in Refs.~\refcite{Avron1995,Levay1995}, and later extended in Refs.~\refcite{Read2009,Read2011} to fractional Hall systems and chiral and anyon superfluids (see also Refs.~\refcite{Tokatly2007,TokatlyErratum2009,Tokatly2009,Haldane2009}). 

The usual approach is to compute the value of the Hall viscosity by deforming adiabatically the geometry. Small geometric deformations $\xi_i$ can be associated to small changes of the metric,
\begin{equation}
g_{ij}=\delta_{ij}+h_{ij}, \ \ h_{ij}=\partial_i\xi_j+\partial_i\xi_j. 
\end{equation}  
Then, to linear order, the stress tensor receives a contribution proportional to the Hall viscosity
\begin{equation}\label{stressten}
T^{ij}\supset \frac{\eta_H}{2} (\epsilon^{ik}\dot{h}_{k}^{j}+\epsilon^{jk}\dot{h}_{k}^{i}).
\end{equation}
In the quantum theory the stress tensor is computed from variations of the energy with respect to the metric
\begin{equation}
T_{ij} =2\left\langle \frac{\partial H}{\partial g^{ij}}\right\rangle.
\end{equation}
Using the adiabatic approximation, the variation of the Hamiltonian has two contributions to leading order
\begin{equation}
\left\langle \frac{\partial H}{\partial g^{ij}}\right\rangle= \frac{\partial E}{\partial g^{ij}}-\Omega_{ijkl}\dot{h}^{kl}.
\end{equation}
The first is the change of the energy of the ground state under the deformation, the elastic modulus tensor is obtained simply by picking the linear term in the metric
\begin{equation}
V \lambda_{ijkl}=-4\frac{\partial^2 E}{\partial g^{ij} \partial g^{kl}}.
\end{equation}
The second term is the adiabatic or Berry curvature, 
\begin{equation}
\Omega_{ijkl}=i\left[  \frac{\partial}{\partial g^{ij}} \left\langle\psi \Big| \frac{\partial}{\partial g^{kl}}\psi\right\rangle-\frac{\partial}{\partial g^{kl}}\left\langle \frac{\partial}{\partial g^{ij}}\psi \Big| \psi\right\rangle\right].
\end{equation}
that is non-trivial if the phase of the state $|\psi\rangle$ changes along a closed path in the space of deformations. In Ref.~\refcite{Read2011} it was argued that the non-zero Hall viscosity is associated to orbital (or 'intrinsic') angular momentum. If we do two shear deformations that preserve the area, the final state differs from the original one by a rotation. If the wavefunction is an eigenstate of angular momentum, the wavefunction gets multiplied by a phase that depends on the spin of the state. 

The result of the adiabatic calculation was that the Hall viscosity is proportional to the average density of particles and an average orbital angular momentum $\bar{s}$ per particle (for instance, for a chiral $p$-wave superfluid $\bar{s}=1/2$)
\begin{equation}\label{Hallviscspin}
\eta_H = \frac{1}{2}\bar{s}\bar{n}.
\end{equation}
For Quantum Hall systems the average density is determined by the filling fraction $\nu$ and the magnetic length $\ell_B=1/\sqrt{B}$
\begin{equation}
\bar{n}=\frac{\nu}{2\pi \ell_B^2}.
\end{equation}
In known examples the orbital angular momentum is related to a topological quantity, the shift ${\cal S}=2\bar{s}$. The shift determines the number of particles $N$ in terms of the number of flux quanta $N_\phi$ when the system is put on a sphere\cite{Wen1992} \footnote{For a system like the chiral superfluid $\nu^{-1}=0$.}
\begin{equation}
N_\phi=\nu^{-1} N-{\cal S}.
\end{equation}
For rational values of $\nu$, this implies that ${\cal S}$ is a rational number as well.  We show the values of the filling fraction and the shift for several topological states in Table~\ref{table1}.

\captionsetup{width=.75\textwidth}
\begin{table}[h]
\begin{center}
\begin{tabular}{|c|c|c|}
\hline  & ${ \nu}$ & ${\cal S}$ \\  \hline
($p_x+ i p_y$) chiral superfluid  & - & $1/2$ \\ \hline
($d_{x^2-y^2}+id_{xy}$) chiral superfluid  & - & $1$ \\ \hline
${\cal N}$th Landau level filled & $1$ & $2{\cal N}+1$ \\ \hline
$N$ first Landau levels filled & $N$ & $N$ \\ \hline
Laughlin & $1/q$ & $q$ \\ \hline
Read-Rezayi\cite{Read1999} & $k/(Mk+2)$ & $M+2$ \\ \hline
\end{tabular}
\caption{\small Values of the filling fraction $\nu$ and shift ${\cal S}$ for different topological states. ${\cal N}$, $N$, $q$, $k$, $M$ are all positive integers.} \label{table1}
\end{center}
\end{table}

If rotational and translational invariance are not broken the shift cannot change continuously, hence it is topologically protected. Otherwise there is no fixed relation between $N_\phi$ and $N$ one can use, although in Ref.~\refcite{Haldane2011} it has been argued that translational invariance is enough to determine the Hall viscosity. 

\subsection{Torsional Hall viscosity and  Chern insulators}\label{sec:tors}

Compared to systems with Galilean invariance, calculations of Hall viscosity at zero temperature in relativistic systems are scarce. An example studied in Refs.~\refcite{Hughes2011,Hughes2013} is a Chern insulator modelled as a free massive Dirac fermion. 

The generalization of \eqref{stressten} to a relativistic system is a term in the {\em symmetric} energy-momentum tensor
\begin{equation}
T^{\mu\nu}\supset -\frac{\eta_H}{2}( \epsilon^{\mu\alpha\beta}\partial_\alpha h_{\beta}^{\nu}+\epsilon^{\nu\alpha\beta}\partial_\alpha h_{\beta}^{\mu}).
\end{equation}
However, this term vanishes for a Dirac fermion. Coupling a Dirac fermion to the metric produces parity-breaking terms that are at least third order in derivatives. \cite{Deser2000} Since the Hall viscosity is of first order, it is zero for this system. It is important to emphasize that the definition of $\eta_H$ given above refers specifically to the symmetric tensor.

There is however a related quantity which enters as a `Hall viscosity' in the canonical energy-momentum tensor. We will dub it as `torsional Hall viscosity' for reasons that will be clear below. For a small change of the metric,
\begin{equation}
g_{\mu\nu}=\eta_{\mu\nu}+h_{\mu\nu},
\end{equation}
the canonical energy-momentum tensor picks up a term of the form 
\begin{equation}
T^\mu_{{\rm can}\ \nu} \supset \frac{\zeta_H}{2} \epsilon^{\mu\alpha\beta}\partial_\alpha h_{\beta\nu}.
\end{equation}
The symmetrized energy-momentum tensor can be obtained from the canonical one by adding an improvement term
\begin{equation}\label{symmtens}
T^\mu_{\ \nu}=T^\mu_{{\rm can}\ \nu}+\nabla_\alpha \Psi^{\alpha\mu}_{\ \ \nu}, \ \ \Psi^{\alpha\mu}_{\ \ \nu}= -\Psi^{\mu\alpha}_{\ \ \nu} .
\end{equation}
The improvement term is defined in terms of the spin current $S^\alpha_{ \ \mu\nu}$ as
\begin{equation}
\Psi^\alpha_{\ \mu\nu}=- S^{\alpha}_{\ \mu \nu}+S_{\mu\ \nu}^{\ \alpha}+S_{\nu\ \mu}^{\ \alpha}.
\end{equation}
There is a term related to the torsional Hall viscosity that contributes to the spin current
\begin{equation}
S^{\alpha}_{\ \mu \nu} \supset -\frac{\zeta_H}{4}\left(\epsilon^{\alpha\beta}_{\ \ \nu}h_{\mu\beta} -\epsilon^{\alpha\beta}_{\ \ \mu}h_{\nu\beta}\right).
\end{equation}
This cancels exactly the torsional Hall viscosity term in the canonical energy-momentum tensor when we compute the symmetrized tensor. A similar observation was made in Ref.~\refcite{Haehl:2013kra} for the effective action of non-dissipative fluids. The original definition of the Hall viscosity $\eta_H$ refers to the symmetric energy-momentum tensor, so strictly speaking the torsional Hall viscosity is different from the usual Hall viscosity $\zeta_H\neq \eta_H$ (for the Dirac fermions $\eta_H=0$ and $\zeta_H\neq 0$). 

Even though there is no contribution to the symmetric tensor, it was proposed in Refs.~\refcite{Hughes2011,Hughes2013} that the canonical energy-momentum tensor can be interpreted as a `stress current' in a generalized theory of elasticity.\cite{Eringen1967,Hehl:2007bn} The fact that the stress tensor and the spin current are not independent can be seen from the point of view of a condensed matter system as a consequence of spin-orbit coupling. A related approach is the calculation in lattice models of the effective phonon action.\cite{Phonon} Coupled to a fluid with a non-zero Hall viscosity, the low-energy effective action for the displacement field $\xi_i$ receives a contribution of the form
\begin{equation}
S\supset \int d^3 x \,\eta^{ijkl}\partial_i \xi_j \partial_k \dot{\xi}_l.
\end{equation}
The coefficient $\eta^{ijkl}$ computed using the adiabatic approximation is proportional to \eqref{hallvisc}, but it also depends on parameters of the lattice. It was suggested in Ref.~\refcite{Phonon} that the `phonon Hall viscosity' could be measured in X-ray diffraction experiments.

Coming back to the Chern insulator, the calculation of the stress current was made by coupling the fermions to a background frame field (vielbein) $e_\mu^{ \ A}$, $A=0,1,2$ and spin connection $\omega_\mu^{\ AB}$.\footnote{The orthogonal frame formulation of general relativity can be found in several books, e.g. Refs.~\refcite{Weinberg,Nakahara,Ortin}}  The physical interpretation is that in a crystalline material the vielbein determines the geometry of the unit cell at a lattice site, while the spin connection also carries information about dislocations in the lattice ({\em torsion}). The coupling of the Dirac fermion to these fields is
\begin{equation}
S=\int d^3 x\, |e|\bar{\psi} \left(e_A^{\ \mu}\gamma^A{\cal D}_\mu -m\right) \psi,
\end{equation}
where $e_\mu^{\ A}$ is the inverse vielbein, $|e|$ the determinant, and the covariant derivative is defined as
\begin{equation}\label{covdevDirac}
{\cal D}_\mu \psi = \left(\partial_\mu-\frac{1}{4}\omega_\mu^{AB}\gamma_{AB} \right)\psi , \ \ \gamma^{AB}=\frac{1}{2}\left[\gamma^A,\gamma^B\right].
\end{equation}
One can then determine the Hall viscosity coefficient by a generalization of the Berry curvature calculation or by a one-loop calculation of the two-point function of the stress current. The result is divergent, so a proper regularization needs to be introduced. The finite result turns out to be\cite{Hughes2011,Hughes2013}
\begin{equation}\label{torsvisc}
\zeta_H=q_T\frac{m^2}{2\pi}\frac{1-{\rm sign}(m)}{2},
\end{equation}
where $q_T$ is the coupling of the fermion to the torsion field (hidden in the spin connection in \eqref{covdevDirac}).  Because of the divergences that appear in the calculation, in Ref.~\refcite{Hughes2013} it is suggested that one should see this result as the {\em difference} between the Hall viscosities of the normal and topological insulator, while the total value depends on the details of the microscopic theory and is not universal. An universal term does appear at higher order, shifting the value of \eqref{torsvisc} by a term proportional to the curvature\cite{Hughes2013}
\begin{equation}\label{Rcorr}
\zeta_H\to \zeta_H+\frac{1}{48\pi}\frac{1-{\rm sign}(m)}{2} R.
\end{equation}

The difference in the value of $\zeta_H$ in different phases produces an anomaly in the effective 1+1 theory at the boundary between them. Physically, the anomaly produces a spectral flow where the energy of the states gets rescaled according to their momentum (zero momentum states are fixed points of the spectrum). The spectral flow can be achieved for instance if the system has the geometry of a cylinder and the radius of the cylinder is changed adiabatically by introducing dislocations. Note that the difference with the spectral flow driven by the chiral anomaly, where the spectrum is shifted uniformly and there are no fixed points.

A related model to the relativistic Chern insulator where both $\zeta_H\neq 0$ and $\eta_H\neq 0$ is the anisotropic (nematic) Chern insulator studied in Ref.~\refcite{You2013}. The model consists of interacting fermions with non-relativistic $d$-wave symmetric dispersion relation in 2+1 dimensions. The system can be in an anisotropic gapped phase characterized by a nematic order parameter $Q_{\mu\nu}$, with $Q_{0\mu}=0$ and $\eta^{\mu\nu}Q_{\mu\nu}=0$. The effective action after integrating out the fermions depends on the effective metric
\begin{equation}
g_{\mu\nu}=\eta_{\mu\nu}+\frac{1}{2m}Q_{\mu\nu}.
\end{equation}
Where $m$ is the mass of the fermions in the gapped phase. The effective action contains a term of the form
\begin{equation}
{\cal L}_{\rm eff}=-\frac{m}{16\pi}\delta^{kl}\epsilon^{ij}g_{ik}\partial_0 g_{jl},
\end{equation}
which introduces a Hall viscosity term for the effective metric. The coefficient is obtained from a Berry phase calculation. The effects of torsion were also studied by coupling the fermions to background vielbeins. It was found that there is a non-universal (cutoff dependent) torsional Hall viscosity $\zeta_H$. A similar term proportional to the Hall viscosity appears in the nematic Hall fluid of Ref.~\refcite{Maciejko2013}. Both models are non-relativistic, we will discuss later some examples of relativistic models where $\eta_H\neq 0$, but those will not correspond to topological phases.

\section{Kubo formulas in Galilean invariant systems}  \label{sec:kub}

The viscosity can also be derived using linear response theory. The coefficients that determine the response of the system to external sources are obtained from correlation functions of the system in the absence of sources. The relation between the coefficient and the correlator is the Kubo formula. A careful derivation of the Kubo formulas for conductivities and viscosities in systems with Galilean invariance was made in Ref.~\refcite{Bradlyn2012}. A Kubo formula for the Hall viscosity in a relativistic theory was given in Ref.~\refcite{Saremi:2011ab}.

Let us review the main steps. We introduce an external gauge field $A_\mu$ that couples to the number current and density and an external metric $g_{ij}$. The current and stress tensor in the presence of general sources are obtained from the partition function ${\cal Z}[A_i,g_{ij}]$ (or the Hamiltonian) by taking a variation with respect to the sources
\begin{equation}
J^\mu=\frac{\partial}{\partial A_\mu}\log{\cal Z}, \ \ T_{ij}=\frac{2}{\sqrt{g}}\frac{\delta}{\delta g^{ij}}\log{\cal Z}.
\end{equation}
Expanding around $A_\mu=0$ and $g_{ij}=\delta_{ij}$, the current and the energy-momentum tensor receive contributions of the form
\begin{align}
&J_i = \sigma_{ij}E_j=-\frac{\bar{n}}{m} A_i + \sigma^0_{ij} E_j+\cdots, \\ &T_{ij}=p\delta_{ij}-\frac{1}{2}\kappa^{-1}\delta_{ij} \delta_{kl}h_{kl} -\frac{1}{2}\eta_{ijkl}\dot{h}_{kl}+\cdots,
\end{align}
where $E_i=\partial_t A_i-\partial_i A_t$ and $\sigma^0$ is the conductivity minus the diamagnetic term. The dots denote possible higher derivative terms. The first term in the current is the diamagnetic current, a contact term in the correlator that can be seen as coming from quadratic terms $\sim A_i^2$ in the action. The term proportional to the inverse compressibility in the stress tensor has a similar origin. We have assumed that the ground state in the absence of sources is isotropic and homogeneous. Other than that, the derivation is very generic, and can be applied to systems at nonzero temperature and density. We can identify the coefficients of the sources in the current and stress tensor with the conductivity and the viscosity. We can then obtain the conductivity and viscosity by taking a second variation with respect to the sources. This relates them to the retarded two-point functions of the current and the stress tensor. We will give the Kubo formula for the conductivity in frequency and momentum space (by considering the conductivity at non-zero momentum we are taking into account higher derivative terms in the expansion of the current):
\begin{equation}\label{condgal}
\sigma_{ij}(\omega,\mathbf{q})=-\frac{i\bar{n}}{m \omega^+}\delta_{ij}+\frac{i}{\omega^+}G^R_{ij}(\omega,\mathbf{q}).
\end{equation}
Where $\omega^+=\omega+i\epsilon$, with $\epsilon\to 0^+$ corresponding to a retarded response.  The Fourier transform of the retarded correlator of the current is defined as
\begin{equation}
G^R_{ij}(\omega,q)=\int dt\int d^d x\, e^{i\omega^+t-i\mathbf{q}\cdot\mathbf{x}}\, i\Theta(t)\left\langle\left[J_i(t,\mathbf{x}) ,\,J_j(0,\mathbf{0})\right] \right\rangle.
\end{equation}
Due to the diamagnetic term, the real part of the conductivity has a delta-function singularity at zero frequency since
\begin{equation}
\frac{1}{\omega^+}={\cal P}\frac{1}{\omega}-i\pi\delta(\omega),
\end{equation}
where ${\cal P}$ is the principal part. This may be avoided if translation invariance is broken for instance by scattering with impurities.

The Kubo formula for the viscosity is
\begin{equation}
\eta_{ijkl}(\omega,\mathbf{q})=\frac{i}{\omega^+}G^R_{ijkl}(\omega,\mathbf{q}),
\end{equation}
where the Fourier transform of the retarded correlator of the stress tensor is defined as
\begin{equation}
G^R_{ijkl}(\omega,\mathbf{q})=\int dt\int d^d x\, e^{i\omega^+t-i\mathbf{q}\cdot\mathbf{x}}\, i\Theta(t)\left\langle\left[T_{ij}(t,\mathbf{x}) ,\,T_{kl}(0,\mathbf{0})\right] \right\rangle.
\end{equation}
Note that the full response $\chi_{ijkl}$ obtained from the second variation of the partition function includes a contact term proportional to the inverse compressibility
\begin{equation}\label{eqchi}
\chi_{ijkl}=\frac{i\kappa^{-1}}{\omega^+}\delta^{ij}\delta^{kl}+\eta^{ijkl}.
\end{equation}
A quantum Hall fluid is incompressible under changes of volume at fixed magnetic field and particle number, but we are interested in the response of the system at fixed {\em filling fraction} rather than magnetic field. Given the relation between the filling fraction and the magnetic field
\begin{equation}
\nu =\frac{2\pi N}{BV},
\end{equation}
if we make a dilatation that changes the volume by a factor $\lambda$, the change in the magnetic field should compensate the dilatation
\begin{equation}
V\to \lambda V, \ \ B\to B/\lambda.
\end{equation}
Therefore,
\begin{equation}\label{volB}
\frac{dV}{V}=-\frac{dB}{B}.
\end{equation}
The total energy $E=V \varepsilon(\nu,B)$ is related to the pressure through the thermodynamic relation $dE=-p dV$. Dividing by the volume and using  \eqref{volB} we find, 
\begin{equation}
p(\nu,B)=B\left(\frac{\partial \varepsilon}{\partial B}\right)_\nu-\varepsilon.
\end{equation}
Therefore, `internal' compressibility is
\begin{equation}\label{comprHall}
\kappa^{-1}(\nu,B)=-V\left(\frac{\partial p}{\partial V}\right)_\nu = B\left(\frac{\partial p}{\partial B}\right)_\nu=B^2\left(\frac{\partial^2 \varepsilon}{\partial B^2}\right)_\nu.
\end{equation}

\subsection{Galilean invariance, viscosity and conductivity}

In a Galilean invariant system where all particles have the same charge and mass, the momentum density $T^{ti}$ and the current are proportional to each other
\begin{equation}\label{galrel}
T^{ti}=m J^i.
\end{equation}
This can be used to derive some relations between the conductivity and the viscosity, and in particular between their Hall components. The original derivation was made in Ref.~\refcite{Hoyos2012} using symmetries of the effective action. A more complete derivation from the microscopic point of view that keeps all possible contributions was made in Ref.~\refcite{Bradlyn2012}. Another derivation for fermions in Landau levels was given in Ref.~\refcite{Biswas2014}. It was shown there that the contribution proportional to the Hall viscosity can be understood semiclassically as produced by shear deformations of cyclotron orbits.
Here we will follow a more heuristic derivation and focus only on the parity breaking components.

The momentum conservation equation in the presence of an external magnetic field $F_{ik}=\partial_i A_k-\partial_k A_i$ is (we set the charge equals to one)
\begin{equation}
\partial_t T^{ti}+\partial_k T^{ki}=F^i_{\ t}J^t+F^i_{\ k} J^k.
\end{equation}
Together with \eqref{galrel}, the stress tensor and the current are related by the equation
\begin{equation}
\partial_k T^{ki}=\left(F^i_{\ t}J^t+F^i_{\ k}-m\delta^i_k\partial_t\right)J^k.
\end{equation}
We will now use that $F_{ik}=B\epsilon_{ik}$, $F^i_{\ t}=-E^i$  and the definition of the cyclotron frequency $\omega_c=B/m$ to write the Fourier transformed equation as
\begin{equation}\label{wardkohn}
q_k T_{ki} =i E_i J^t+m \left(\omega \delta_{ik}-i\omega_c\epsilon_{i k}\right)J_k.
\end{equation} 
The inverse of this relation is
\begin{equation}\label{kohn}
J_i=\frac{1}{m}\frac{\omega \delta_{il}+i\omega_c\epsilon_{il}}{\omega^2-\omega_c^2}(-iE_l J^t+q_k T_{kl}),
\end{equation}
which is Kohn's theorem:\cite{Kohn1961} the motion of the center of mass of the fluid has a cyclotron resonance at $\omega=\omega_c$ which is protected by Galilean invariance. 

At the level of correlation functions \eqref{wardkohn} is manifested as a set of Ward identities. A Ward identity for the correlator between the current and the stress tensor can be found by taking a variation of \eqref{wardkohn} with respect to the gauge field $A_l$ and setting the sources to zero.
\begin{equation}
q_k \vev{T_{ki} J_l}+\frac{\delta p}{\delta A_l}q_i =i\bar{n}\omega_c \epsilon_{il}+m(\omega \delta_{ik} -i\omega_c \epsilon_{ik}) \vev{J_k J_l}.
\end{equation}
Where we have used that $\vev{J^t}=\bar{n}$, $\vev{T_{ij}}=p\delta_{ij}$ when the sources are set to zero. The variation of the pressure gives a term proportional to the inverse compressibility given in \eqref{comprHall}
\begin{equation}
\frac{\delta p}{\delta A_i}=\frac{\kappa^{-1}}{B}i\epsilon_{ki}q_k.
\end{equation}
A similar Ward identity exists for the stress-stress correlator, if we take a variation of \eqref{kohn} with respect to the inverse metric $g^{ni}$
\begin{align}
\frac{1}{2}q_k q_n\vev{T_{kj} T_{ni}} +\frac{\delta p}{\delta g^{ni}}q_jq_n+p q^2\delta_{ij}=m(\omega \delta_{jl}-i\omega_c \epsilon_{jl})q_n \vev{J_l T_{ni}}.
\end{align}
Combining the two identities we find
\begin{align}\label{ward2}
\notag \frac{1}{2}q_k q_n\vev{T_{kj} T_{ni}} & +\frac{\delta p}{\delta g^{ni}}q_jq_n+p q^2\delta_{ij}\\ &=m(\omega \delta_{jl}-i\omega_c \epsilon_{jl})\left[-\frac{\kappa^{-1}}{B}i\epsilon_{nl}q_n q^i -i\bar{n}\omega_c \epsilon_{il}+m(\omega \delta_{ik} +i\omega_c \epsilon_{ik}) \vev{J_l J_k}\right].
\end{align}
The correlator of stress tensors is proportional to the viscosity,
\begin{equation}
\vev{T_{kj} T_{ni}}=-i\omega \eta_{kjni},
\end{equation}
 while the correlator of the currents is proportional to the conductivity
\begin{equation}
\vev{J_l J_k}=-i\omega \sigma_{lk}+\frac{\bar{n}}{m}\delta_{lk}.
\end{equation}
Multiplying \eqref{ward2} by $i/(2\omega)$ and contracting with $\epsilon_{ji}$,
\begin{equation}
\frac{1}{4}\epsilon_{ji}q_n q_k \eta_{kjni} =m\omega_c\bar{n}   -\frac{\kappa^{-1}}{2\omega_c} q^2-(m\omega_c)^2 \sigma_H+O(\omega,q^4),
\end{equation}
where we have defined
\begin{equation}
\sigma_H=\frac{1}{2}\epsilon_{ji}\sigma^{ij}.
\end{equation}
Then, solving for $\sigma_H$ and using the expression for the Hall viscosity \eqref{hallvisc},
\begin{equation}\label{condvsvisc}
\sigma_H=\frac{\bar{n}}{m\omega_c}+\frac{q^2}{2(m\omega_c)^2}\left[\eta_H-\frac{\kappa^{-1}}{\omega_c} \right]+O(q^4).
\end{equation}
The first term is the usual Hall conductivity, while the second term is an $O(q^2)$ correction depending on the Hall viscosity and the inverse compressibility. An interesting consequence is that one could then set an experiment where inhomogeneous electric fields are used to measure the Hall viscosity from the generated current, in principle an easier task than measuring it through stresses.

In a system where there is no magnetic field, the Kubo formula is changed to\cite{Bradlyn2012}
\begin{equation}\label{condvsvisc2}
\sigma_H(\omega)=\frac{q^2}{m\omega^2}\eta_H(\omega)+O(q^4).
\end{equation}
In particular, the static Hall conductivity vanishes at zero momentum.

\section{Hall viscosity in effective theories} \label{sec:hyd}

Effective theories are useful descriptions at low energies. They capture the relevant dynamics and symmetries of the system in a simpler fashion than in the full microscopic theory. Corrections to the effective theory can be computed systematically up to some energy scale where the effective description breaks down. If parity and time reversal invariance are broken, we expect the Hall viscosity to be captured by the effective description. In a very rough way we can classify those systems according to their spectrum:
\begin{itemize}
\item {\it Fully gapped systems}: such as integer and fractional Hall states, that can be viewed as incompressible fluids. For these systems one can integrate out completely all the degrees of freedom, and work with a local generating functional depending on the external sources. Other possible candidates are topological insulators with broken parity and time reversal invariance.
\item {\it Partially gapped systems}: as for instance chiral or anyon superfluids, where a few degrees of freedom (the Goldstone bosons) remain massless. In order to correctly capture the low energy behaviour of the system one should include the massless degrees of freedom in the effective description, coupled to external sources.
\item {\it Gapless systems}: generically fluids at finite temperature. The effective theory is hydrodynamics, but since the Hall viscosity is dissipationless it should be possible to capture it with a theory of dissipationless fluids. In this case we would expect that there is an effective action for the fluid that captures the Hall viscosity.
\end{itemize}

\subsection{Sources and symmetry transformations}\label{transf}

The common strategy is to find the dependence of the effective action on external sources for the current and the energy-momentum tensor. These are background gauge fields $A_\mu$ and metric $g_{\mu\nu}$. The current and energy-momentum tensor are computed from the partition function ${\cal Z}\left[A_\mu, g_{\mu\nu}\right]$ by taking variations with respect to the sources
\begin{equation}
J^\mu =\frac{\delta }{\delta A_\mu}\log {\cal Z}, \ \ T^{\mu\nu} = -\frac{2}{\sqrt{-g}}\frac{\delta}{\delta g_{\mu\nu}}\log {\cal Z}.
\end{equation}
In the presence of external sources the global symmetries of the theory can be enhanced to local symmetries under which the partition function is invariant.
This imposes strong constraints in the form of the effective action.

Instead of using the metric as external source, it is sometimes more convenient to work with the vielbeins $e_\mu^{\ A}$, $A=0,1,2$ and the spin connection $\omega_\mu^{AB}$ that we introduced in \S~\ref{sec:tors}.  In non-relativistic theories we will restrict to spatial vielbeins $e_i^{\ a}$, $a=1,2$ and metric. Geometrically the vielbeins are local changes of basis between the coordinate frame and an orthogonal frame. The spin connection determines how the orthogonal frame rotates under parallel transport. As was mentioned before a possible interpretation in a crystalline solid is that the vielbeins describe the geometry of the unit lattice at different lattice sites.

The expressions of the metric and spin connection in terms of the vielbeins are
\begin{eqnarray}
&{\rm relativistic} &g_{\mu\nu}=\eta_{AB}e_\mu^{\ A} e_\nu^{ \ B},  \ \ \omega_\mu^{AB}=\eta^{AC}e_C^{\ \nu}\nabla_\mu e_\nu^{ \ B},\\
\label{spincon} &{\rm non-relativistic} &g_{ij}=\delta_{ab} e_i^{\ a} e_j^{\ b},  \ \ \omega_\mu=\frac{1}{2}\epsilon_{ab}e^{a\, i}\nabla_\mu e_i^{ \ b}.
\end{eqnarray}
The covariant derivative $\nabla_\mu$ is defined with the connection $\Gamma$,
\begin{equation}
\nabla_\mu e_\nu^{\ A} =\partial_\mu e_\nu^{\ A}-\Gamma_{\mu\nu}^\sigma e_\sigma^{\ A}.
\end{equation}
We will usually take $\Gamma$ to be the Levi-Civita connection, but in some cases we will allow a torsion component $\Gamma_{\mu\nu}^\sigma-\Gamma_{\nu\mu}^\sigma\neq 0$ as well. Note that, under local $O(2)$ rotations $\delta e_i^{\ a}=\lambda \epsilon^a_{\ b}e_i^{\ b}$, $\omega_\mu$ transforms like an
Abelian gauge potential $\omega_\mu \to \omega_\mu -\partial_\mu \lambda$. In fact we can interpret the spin connection as the gauge field for rotations of the orthogonal frame indices.

A theory coupled to the gauge field and metric should be invariant under gauge transformations $\alpha$ and diffeomorphisms $\xi^\mu$. Under an infinitesimal transformation the sources change as
\begin{align}
&\delta A_\mu=-\partial_\mu \lambda-{\cal L}_\xi A_\mu =-\partial_\mu \lambda-\xi^\alpha\partial_\alpha A_\mu-\partial_\mu \xi^\alpha A_\alpha,\\
&\delta g_{\mu\nu} =-{\cal L}_\xi g_{\mu\nu}= -\xi^\alpha\partial_\alpha g_{\mu\nu}-\partial_\mu\xi^\alpha g_{\alpha\nu}-\partial_\nu \xi^\alpha g_{\mu\alpha}.
\end{align}
If we work with vielbeins and spin connection, we  also should take into account local frame rotations $\lambda^A_{\ B}$. The transformation of the sources is
\begin{align}
\delta e_\mu^{\ A} &=\lambda^A_{\ B}e_\mu^{\ B}-{\cal L}_\xi e_\mu^{\ A} =\lambda^A_{\ B}e_\mu^{\ B}-\xi^\alpha\partial_\alpha e_\mu^{\ A}-\partial_\mu \xi^\alpha e_\alpha^{\ A},\\
\notag \delta \omega_\mu^{AB} &=-{\cal D}_\mu \lambda^{AB} -{\cal L}_\xi \omega_\mu^{AB}=\\
&
-\partial_\mu\lambda^{AB}+\lambda^A_{ \ C}\omega_\mu^{CB}+\lambda^B_{ \ C}\omega_\mu^{AC}-\xi^\alpha\partial_\alpha\omega_\mu^{AB}- \partial_\mu\xi^\alpha\omega_\alpha^{AB}.
\end{align}
So far these transformations are valid for relativistic systems. Their non-relativistic counterparts were introduced in Ref.~\refcite{Son:2005rv} to constrain the effective theory of unitary Fermi gases.\cite{FermiGas} Later they were extended for Hall systems in Ref.~\refcite{Hoyos2012}. In Ref.~\refcite{Son2013} they were formulated in the context of Newton-Cartan theory and extended to take into account a non-zero electron spin $g$-factor. The transformation of the gauge field and spatial metric are
\begin{align}
  \delta A_t &=-\dot{\lambda} -\xi^k \partial_k A_t - A_k \dot\xi^k+\frac{\mathfrak{g}}{4}\varepsilon^{ij}\partial_i(g_{jk}\dot{\xi}^k), \label{GCT-A0} \\
  \delta A_i &= -\partial_i\lambda -\xi^k \partial_k A_i - A_k \partial_i \xi^k - m g_{ik}\dot \xi^k ,
    \label{GCT-Ai}\\
  \delta g_{ij} &= -\xi^k \partial_k g_{ij} - g_{kj} \partial_i \xi^k 
                   - g_{ik}\partial_j \xi^k . \label{GCT-gij} 
\end{align}
Where $\varepsilon^{ij}=\epsilon^{ij}/\sqrt{g}$, $\mathfrak{g}$ is the spin $g$-factor ($\mathfrak{g}\simeq 2$ for electrons in vacuum) and $m$ is the mass of the particles. Note that the gauge field changes inhomogeneously under Galilean boosts $\xi^i=v^i t$. This is to compensate the change in the electronic wavefunction, that transforms as a projective representation of the Galilean group 
\begin{equation}
\Psi(t,\mathbf{x}) \to e^{-im \mathbf{v}\cdot \mathbf{x}+im \frac{v^2}{2}t} \Psi(t,\mathbf{x}-\mathbf{v} t).
\end{equation}
The transformations of the spatial vielbeins and non-relativistic spin connection are
\begin{align}
&\delta e_i^{\ a}=\lambda\epsilon^a_{\ b}e_\mu^{\ b}-{\cal L}_\xi e_i^{\ a} =\lambda\epsilon^a_{\ b}e_i^{\ b}-\xi^k\partial_k e_i^{\ a}-\partial_i \xi^k e_k^{\ a},\\
&\delta \omega_t =-\dot{\lambda} -\xi^k\partial_k \omega_t -\dot{\xi}^k\omega_k-\varepsilon^{ij}\partial_i(g_{jk}\dot{\xi}^k), \\
&\delta \omega_i =-\partial_i\lambda -{\cal L}_\xi \omega_i=
-\partial_i\lambda-\xi^k\partial_k\omega_i- \partial_i\xi^k\omega_k.
\end{align}
The inhomogeneous transformations of the $A_t$ and $\omega_t$ components of the gauge field can be compensated with the help of the magnetic field density $B=\varepsilon^{ij}\partial_i A_j$ ($\varepsilon^{ij}=\epsilon^{ij}/\sqrt{g}$). We define the corrected gauge potential and spin connection as
\begin{align}\label{altpots}
&\tilde{A}_t=A_t+\frac{\mathfrak{g}}{4}B,\\
&\tilde{\omega}_t=\omega_t-\frac{B}{m}.
\end{align}
In the Newton-Cartan formalism it is also possible to correct the gauge potential $A_i$ with the help of a velocity vector.\cite{Son2013}

\subsection{Integer and fractional Hall states}

We will start with quantum Hall states of systems with Galilean invariance
and made up of particles of the same charge/mass ratio. Quantum Hall states have an effective hydrodynamic description as incompressible fluids (for constant magnetic field).\cite{Abanov2012,Wiegmann2012,Son2013} As they are gapped states, by integrating out the velocity field one can capture the response to external sources with a local functional. We will construct this local functional by employing the formalism of nonrelativistic diffeomorphism invariance.\cite{Son:2005rv,Hoyos2012} A generalization of this analysis to relativistic theories was made very recently in Ref.~\refcite{Son2014}.

To organize a derivative expansion, one needs a power-counting scheme with a small parameter.  There is an ambiguity in choosing the scheme,
as the time derivative $\partial_t$ and spatial derivatives can be chosen to
be independent expansion parameters.  For definiteness, we use the following scheme.  All quantities will be regarded as
proportional to some powers of a small parameter $\epsilon$, times
some powers of $\omega_c$ and $\ell_B$.  The external fields are assumed
to vary slowly in space and time,
\begin{equation}
  \partial_i \sim \epsilon\ell_B^{-1}, \quad \partial_t \sim \epsilon^2 \omega_c .
\end{equation}
As for the magnitude of external perturbations, we assume
\begin{equation}
  \delta A_t \sim \epsilon^0 \omega_c, \quad
  \delta A_i \sim \epsilon^{-1} \ell_B^{-1},\quad
  \delta g_{ij} \sim 1 .
\end{equation}
In this scheme, we allow for order one variations of the metric, the
magnetic field ($\delta B\sim
\epsilon^0\ell_B^{-2}$) and the chemical potential ($A_t$).  
Two important ingredients in our construction of the effective field
theory are the Chern-Simons action and the Wen-Zee action.  The Chern-Simons
action is
\begin{equation}
  S_{\rm CS} = \frac\nu{4\pi}\!\int\!{\rm d}t\,{\rm d}^2x\, 
       \epsilon^{\mu\nu\lambda} A_\mu\partial_\nu A_\lambda\,,
\end{equation}
and is of order $\epsilon^0$ in our power counting scheme.  This will
be the leading term in the effective action.  To construct the Wen-Zee
action, we use the spin connection. 
By using $\omega_\mu$ we can construct the following gauge invariant
action
\begin{equation}\label{wenzeeact}
  S_{\rm WZ} = \frac{k}{2\pi}\!\int\!{\rm d}t\, {\rm d}^2x\,
     \epsilon^{\mu\nu\lambda}\omega_\mu \partial_\nu A_\lambda .
\end{equation}
This action is of order $\epsilon^2$ in our power counting scheme and
has to be included if we work to that order.  The $\omega d\omega$
Chern-Simons term, on the other hand, is of order $\epsilon^4$ and
will not be considered.

The coefficient $k$ is related to the shift ${\cal S}$.  Indeed,
$\partial_1\omega_2-\partial_2\omega_1=\frac12\sqrt g R$ where $R$ is the
scalar curvature.  Integrating by parts, the Wen-Zee action contains a
term
\begin{equation}
  \frac{k}{2\pi}\epsilon^{\mu\nu\lambda} \omega_\mu \partial_\nu A_\lambda 
  \simeq \frac{k}{4\pi}\sqrt g\, A_t R + \cdots
\end{equation}
which gives a contribution to the particle number density that is
proportional to the scalar curvature.  If the quantum Hall state lives on
a closed two dimensional surface, then the total number of particles is
\begin{equation}\label{QNchi}
  Q= \!\int\!{\rm d}^2x\, \sqrt g\, J^t
   = \!\int\!{\rm d}^2x\, \sqrt g\left( \frac\nu{2\pi} B 
     {+} \frac{k}{4\pi} R\right)
   = \nu N_\phi {+} k \chi
\end{equation}
where $N_\phi$ is the total number of magnetic fluxes and
$\chi$ is the Euler character ($\chi=2$ for a sphere).  Comparing to the definition of
${\cal S}$ in Ref.~\refcite{Wen1992}, we find
$k=\frac12\nu{\cal S}$.

The Wen-Zee action gives rise to Hall viscosity.
Expanding the WZ term to quadratic order, one finds, among other
terms,
\begin{equation}
  S_{\rm WZ} = -\frac{k B}{16\pi}\epsilon^{ij} \delta g_{ik}\partial_t \delta g_{jk}
  + \cdots
\end{equation}
which implies the presence of an odd term in the stress tensor two
point function, or Hall viscosity.  The value of the Hall viscosity is
$\eta_H = k B/4\pi= \frac14 {\cal S} \bar{n}$.  

It is straightforward to verify that both $S_{\rm CS}$ and $S_{\rm
  WZ}$ are not invariant under non-relativistic diffeomorphisms, and need to be corrected.  For simplicity we will restrict to the case $\mathfrak{g}=0$. To order $O(\epsilon^2)$, the most general effective action can
be written as $S=\int\!{\rm d}t\,{\rm d}^2x\,\sqrt g\, \sum_{i=1}^5
{\cal L}_i$, where ${\cal L}_i$ ($i=1,\ldots,5$) are five  independent general
diffeomorphism invariant (to order $\epsilon^2$) terms\cite{Hoyos2012}
\begin{align}
  {\cal L}_1 &= \frac\nu{4\pi}\! 
         \Bigl( \varepsilon^{\mu\nu\lambda} A_\mu \partial_\nu A_\lambda
      +  \frac mB g^{ij} E_i E_j\Bigr),\\
  {\cal L}_2 & = \frac{k}{2\pi}\! 
          \Bigl(\varepsilon^{\mu\nu\lambda}\omega_\mu\partial_\nu A_\lambda
          - \frac1{2B}\, g^{ij} \partial_i B\, E_j\Bigr),\\ 
  {\cal L}_3 &= -
     \varepsilon(B) + \frac mB \varepsilon''(B) g^{ij}\partial_i B\, E_j,\\
  {\cal L}_4 &= -\frac12 
          K(B) g^{ij}\partial_i B\,\partial_j B, \\
  {\cal L}_5 &=  R\, h(B),
\end{align}
where $\varepsilon(B)$, $K(B)$, and $h(B)$ are functions of $B$.  The
function $\varepsilon(B)$ has the physical meaning of the energy density
of the quantum Hall state as a function of the magnetic field $B$,
${\cal L}_4$ and ${\cal L}_5$ do not enter the quantities of of our
interest.  The next to leading order term in ${\cal L}_1$ enforces
compliance with Kohn's theorem.  The two-point function of the
electromagnetic current $J^\mu$ is obtained by taking the second
derivative of the effective action with respect to $A_\mu$, then
setting perturbations to zero.  Equivalently we can differentiate the
effective action once with respect to the external fields to get the
current. We find, in flat space
\begin{equation}
  J^i = \frac\nu{2\pi}\epsilon^{ji}E_j + \frac1B\left[\frac{k}{4\pi}
    - m\varepsilon''(B)\right]\epsilon^{ij}\partial_j({\bf \nabla}\!\cdot\!{\bf E})
    + \cdots 
\end{equation}
where $\cdots$ refers to term that vanish when the magnetic field is
not perturbed. This reproduces the relation between Hall conductivity and viscosity that we discussed previously \eqref{condvsvisc}.

We have constructed the first terms of the effective action using purely symmetry arguments. An actual calculation of the effective action that confirms these results was made in Ref.~\refcite{Abanov2014} by integrating out free non-relativistic fermions in Landau levels in the presence of a background metric and gauge fields. The analysis goes beyond what has been presented here, in Ref.~\refcite{Abanov2014} it was also computed the coefficient of the gravitational Chern-Simons $\omega d\omega$, which contributes to the first curvature correction to the Hall viscosity.

\subsection{Chiral and anyon superfluids}

We now study systems with a few massless degrees of freedom at low energies separated by a gap from other excitations, in particular superfluids with broken time reversal invariance, such as the $p$-wave superfluid state of thin films of $^{3}\text{He-A}$.\cite{Vollhardt,Volovik} Galilean invariant effective actions of chiral superfluids  and anyon superfluids were constructed in Ref.~\refcite{Stone2004} and Refs.~\refcite{Chen1989,Greiter1989}. Their extension to incorporate invariance under non-relativistic diffeomorphisms has been made in Ref.~\refcite{Hoyos2013}.

The low energy effective theory consists of a single gapless Goldstone mode in the spectrum, coupled to external sources. In the case of the chiral superfluid there is a $p$-wave condensate of the form
\begin{equation}
\Delta = (p_x\pm ip_y) \hat\Delta.
\end{equation}
Where $\hat\Delta$ is real. A $SO(2)_V$ rotation would change the value of the condensate by a phase, since
\begin{equation}
(p_x\pm ip_y)\to e^{\pm i\lambda} (p_x\pm ip_y).
\end{equation}
A $U(1)_N$ transformation also changes the value of the condensate by a phase
\begin{equation}
\hat\Delta\to e^{2i\alpha}\hat\Delta.
\end{equation}
Therefore the symmetry group of spatial rotations and particle number is broken to a diagonal symmetry $SO(2)_V\times U(1)_N\to U(1)_D$. Associated to the broken generator there is a Goldstone boson $\theta$. Its transformation under $SO(2)_V$ and $U(1)_N$ symmetries is 
\begin{equation}
\delta\theta= -\alpha\pm\frac{1}{2}\lambda.
\end{equation}
The sign is determined by chirality of the ground state.
Under the spatial diffeomorphisms described in subsection \ref{transf}, the Goldstone boson transforms as a scalar
\begin{equation}
\delta_\xi\theta=-\xi^k\partial_k\theta.
\end{equation}
Without referring explicitly to the condensate, one can generalize the Goldstone field to cases with different orbital angular momentum $2\bar{s}$. The Goldstone field then transforms as
\begin{equation}
\delta\theta= -\xi^k\partial_k\theta-\alpha-\bar{s}\lambda.
\end{equation}
For a chiral superfluid with pairing in the $\ell^{\text{th}}$ partial wave the value is $\bar{s}=\pm \ell/2$.  More general values of $\bar{s}$ are also possible. For anyon superfluids with a fractional statistical phase\cite{Chen1989,Greiter1989}
\begin{equation}
\Theta=\pi\left(1-\frac1\ell\right),
\end{equation}
the value of $\bar{s}$ is
\begin{equation}
\bar{s}=\frac12\left(\ell- \frac{1}{\ell}\right).
\end{equation}
For $\ell>1$ the orbital angular momentum of the anyon superfluid is fractional. When $\ell=1$ the anyon becomes a boson (scalar) and is neutral under $SO(2)_V$ transformations.

In order to construct an invariant action, we define a covariant derivative given by
\begin{equation} \label{covder}
{\cal D}_{\nu} \theta\equiv \partial_{\nu}\theta-\tilde{A}_{\nu}-\bar{s} \tilde{\omega}_{\nu}.
\end{equation}
Where $\theta$ is the Goldstone field, $\tilde{\omega}_t$ and $\tilde{A}_t$ are given in \eqref{altpots} and $\tilde{\omega}_i=\omega_i$, $\tilde{A}_i=A_i$.  With the help of the covariant derivative we can construct the following scalar under non-relativistic diffeomorphisms
\begin{equation} \label{X}
X={\cal D}_t \theta +\frac{g^{ij}}{2} {\cal D}_i \theta{\cal D}_j \theta.
\end{equation}
We have set $m=1$ for simplicity.  Then, the leading order action of the Goldstone field $\theta$  is
\begin{equation} \label{LOS}
S[\theta]=-\int dt d^2 x \sqrt{g} P\left( X\right),
\end{equation}
The superfluid ground state has a finite density, that we characterize by the chemical potential $\mu$. In the effective action it enters as a background value for the Goldstone field, that is decomposed as
\begin{equation} \label{theta}
\theta=\mu t+\varphi
\end{equation}
with $\varphi$ standing for a phonon fluctuation around the ground state. This allows to identify the function $P(X)$ in \eqref{LOS} with the thermodynamic pressure as a function of the chemical potential $\mu$ at zero temperature. 
Time reversal and parity act nontrivially as
\begin{equation}
\begin{split}
T:& \, t\to -t, \, \theta\to -\theta, \, A_i\to -A_i, \, \omega_t\to -\omega_t; \\
P:& \,  x_1\leftrightarrow x_2, \, A_1\leftrightarrow A_2, \, \omega_t\to-\omega_t, \, \omega_1\leftrightarrow -\omega_2.
\end{split}
\end{equation}
For a fixed $\bar{s}\ne 0$, the effective theory \eqref{LOS} is not separately invariant under time reversal $T$  or parity $P$. On the other hand, $PT$ is a symmetry of the theory for any value of $\bar{s}$.

By introducing the superfluid density $\bar{n}\equiv dP/dX$ and the superfluid velocity $v_j\equiv {\cal D}_j \theta$ the nonlinear equation of motion for the Goldstone field can be written in the general covariant form
\begin{equation} \label{EOMgr}
\frac{1}{\sqrt{g}}\partial_t (\sqrt{g} \bar{n})+\nabla_i \left(\bar{n} v^i \right)=0,
\end{equation}
which is the continuity equation in curved space. By linearizing the equation of motion \eqref{EOMgr} in the absence of background gauge fields ($A_{\nu}=\omega_{\nu}=0$) one finds the low-momentum dispersion relation of the Goldstone field to be
\begin{equation}
\omega^2=c_{s}^2 \mathbf{p}^2,
\end{equation}
where the speed of sound $c_s\equiv \sqrt{\partial P/\partial\bar{n}}$ is evaluated in the ground state.

The Hall viscosity can be easily computed from the expansion of the effective action, it contains a term proportional to the spin connection
\begin{equation}
S \supset \int dt d^2 x\sqrt{g}\, \bar{s}\bar{n} \omega_t.
\end{equation}
This takes the same form as the Wen-Zee term in the Hall fluid \eqref{wenzeeact} with the density $\bar{n}$ substituting the magnetic field. From here we can read directly the Hall viscosity in the chiral/anyon superfluid
\begin{equation}
\eta_H=\frac{1}{2}\bar{s}\bar{n}.
\end{equation}
The calculation of the stress tensor in the linear response approximation in Ref.~\refcite{Hoyos2013} shows that this is the full contribution at tree-level, there are no corrections from the fluctuations of the Goldstone modes. New parity-breaking terms do appear at non-zero momentum. The relation between the Hall conductivity and viscosity holds \eqref{condvsvisc2}, but contrary to the Hall viscosity, the Hall conductivity is produced by the propagation of the Goldstone modes:
\begin{equation}
\sigma_H=\frac{\bar{s}\bar{n}}{2}\frac{q^2}{\omega^2-c_s^2 q^2}.
\end{equation}

\subsection{Topological insulators with torsion}

As we saw in \S~\ref{sec:tors}, when a massive Dirac fermion is coupled to a vielbein, there is a response in the canonical energy-momentum tensor that takes a form similar to the Hall viscosity. A local generating functional can be computed by integrating out the massive fermion in the presence of background sources. To leading order in derivatives, the resulting functional takes the form of a Chern-Simons term for the vielbeins\cite{Hughes2011,Hughes2013}
\begin{equation}\label{torsionact}
S_{\rm tors}=\frac{\zeta_H}{2}\int d^3 x \,\eta_{AB}\epsilon^{\mu\nu\lambda}e_\mu^{\ A}{\cal D}_\nu e_\lambda^{\ B}.
\end{equation}
Where the covariant derivative is defined with the spin connection
\begin{equation}
{\cal D}_\mu e_\nu^{ \ A}=\partial_\mu e_\nu^{ \ A}-\omega_{\mu\ B}^A e_\nu^{\ B}.
\end{equation}
There are no other terms at first order in derivatives, other than the usual Chern-Simons term for a gauge field. At higher order there are many more terms, including a Chern-Simons for the spin connection, that produce other interesting effects. We will not comment of those, a thorough analysis can be found in Ref.~\refcite{Hughes2013}.

The Chern-Simons functional \eqref{torsionact} is invariant under all the symmetries, but for the ordinary spin connection it will be exactly zero unless there is a non-zero torsion $\Gamma_{[\mu\nu]}^\sigma=T_{\mu\nu}^\sigma$, since
\begin{equation}
{\cal D}_{[\mu} e_{\nu]}^{\ A}=-T_{\mu\nu}^\sigma e_\sigma^{\ A}.
\end{equation}
Alternatively, one can treat the vielbein and the spin connection as independent fields. In this case the torsion can be zero or non-zero. 

An interesting analogy with the Hall effect is the appearance of edge currents.
The vielbeins can be treated as a collection of gauge fields for translations, charged under the Lorentz group (see e.g. section 4.5 in Ref.~\refcite{Ortin} and section 6 of Ref.~~\refcite{Hughes2013}). The gauge transformation associated to translations is
\begin{equation}
\delta e_\mu^{\ A}=-{\cal D}_\mu\alpha^A.
\end{equation} 
When the curvature associated to the spin connection vanishes, the torsion is invariant under this transformation and the vielbeins can be treated effectively as Abelian gauge fields.

A gauge transformation will induce anomalous terms at the boundaries of space:\footnote{Had we considered diffeomorphism rather than local translation invariance the Ward identity for the current would be modified by terms proportional to the torsion, in such a way that there are no anomalous contributions to the diffeomorphism Ward identity.\cite{Hughes2013} }
\begin{equation}
\delta_\alpha S_{\rm bound} = -\frac{\zeta_H}{2}\oint d^2x \, \alpha_A \epsilon^{mn}T_{mn}^A.
\end{equation}
Where $m,n=0,1$ are the indices after doing the pullback to the boundary. These terms can be interpreted as chiral anomalies of the associated consistent currents
\begin{equation}
{\cal D}_m J_{\rm cons}^{A\,m} =\frac{\zeta_H}{2}\epsilon^{mn}T_{mn}^A.
\end{equation} 
One can then follow Callan-Harvey anomaly inflow argument\cite{Callan1984} to argue that in order to recover the right anomaly there must be edge currents. The reason is that the consistent current does not capture the total transfer of `charge' (energy and momentum) between the bulk and the boundaries. The flow is determined instead by the covariant current
\begin{equation}
J^{A\, m}_{\rm cov} = \zeta_H\epsilon^{mn}e_n^{\ A}.
\end{equation}
Both currents differ by a local term that cannot be obtained from the variation of a local functional.

There are other interesting effects but, as we commented in \S \ref{sec:tors}, there is no Hall viscosity term in the symmetric energy-momentum tensor. The canonical energy-momentum tensor is computed by taking a variation with respect to the vielbein {\em holding the spin connection fixed}.
\begin{equation}
T^\mu_{{\rm can}\ A}= \frac{1}{|e|}\frac{\delta S_{\rm tors}}{\delta e_\mu^{\ A}}=\zeta_H\eta_{AB}\varepsilon^{\mu\nu\lambda} {\cal D}_\nu e_\lambda^{\ B}.
\end{equation}
On the other hand, the spin current is computed by taking the variation with respect to the spin connection
\begin{equation}
S^\mu_{AB}= \frac{\delta S_{\rm tors}}{\delta \omega_\mu^{A B}}=-\frac{\zeta_H}{2}\eta_{AC}\eta_{BD}\epsilon^{\mu\nu\lambda} e_\nu^{\ C} e_\lambda^{\ D}.
\end{equation}
Expanding around $\omega_\mu^{AB}=0$, $e_\mu^{\ A}=\delta_\mu^A+\frac{1}{2} h_\mu^A$,
\begin{align}
&T^\mu_{{\rm can}\ A} =\frac{\zeta_H}{2}\epsilon^{\mu\nu\lambda}\partial_\nu h_{A\lambda},\\
&S^\mu_{AB}=-\frac{\zeta_H}{2}(\epsilon^{\mu\nu}_{ \ \ B}h_{A\nu}-\epsilon^{\mu\nu}_{\ \ A} h_{B\nu}).
\end{align}
As we explained in \S~\ref{sec:tors}, the combination of the canonical energy-momentum tensor and the divergence of the spin current in the symmetric energy-momentum tensor cancels out.


\subsection{Dissipationless hydrodynamics}

We will now discuss gapless systems, in particular fluids with broken parity. We do not expect in this case the Hall viscosity to be quantized in general. For instance, in a classical magnetized plasma the Hall viscosity at high temperatures has the form\cite{Read2009}
\begin{equation}
\eta_H=\frac{T \bar{n}}{2\omega_c}.
\end{equation}
A similarly looking formula can be derived for the torsional Hall viscosity of a relativistic plasma with a background vorticity $\Omega$\cite{Leigh:2012jv} 
\begin{equation}
\zeta_H=\frac{\varepsilon+p}{\Omega}.
\end{equation}
As we saw in \S \ref{sec:def}, the Hall viscosity is non-dissipative. Therefore, in order to study it we can focus on non-dissipative hydrodynamics. In the effective theories of non-relativistic systems we were able to obtain the Hall viscosity directly from local terms depending on the sources in the effective action. In a fluid  there are gapless degrees of freedom, so the generating functional is non-local for general time-dependent configurations. However, since equal-time correlators in field theory decay exponentially at finite temperature, a local generating functional exists for static configurations at thermal equilibrium (assuming there are no other dynamical degrees of freedom such as Goldstone bosons).  Unfortunately this method cannot be applied since the Hall viscosity is proportional to the time derivative of the metric and the shear tensor and vanishes at thermal equilibrium.\cite{Banerjee:2012iz,Jensen:2012jh}\footnote{An explicit computation of the generating functional in a system with free massive fermions was made in Ref.~\refcite{Manes2013}.}

A different approach to describe non-dissipative effects is to construct an effective action where the hydrodynamic degrees of freedom are captured by Goldstone bosons of the broken space symmetries.\cite{Dubovsky:2005xd}\footnote{A review of effective theory descriptions of hydrodynamics is Ref.~\refcite{Jackiw:2004nm}.} In a $d+1$ charged fluid there are $d$ fields $\phi^I(t,\mathbf{x})$ that are the comoving coordinates of the fluid element in the physical space $\mathbf{x}$. The theory is invariant under volume-preserving diffeomorphisms
\begin{equation}
\phi^I\to \xi^I(\phi), \ \ {\rm det}\left(\frac{\partial \xi}{\partial \phi} \right)=1.
\end{equation}
If the theory is not a superfluid, a ``chemical shift'' invariance is also imposed
\begin{equation}
\psi\to \psi+f(\phi).
\end{equation}
This theory possess an exactly conserved current $\nabla_\mu J^\mu=0$
\begin{equation}
J^\mu=\frac{1}{d!}\epsilon^{\mu\alpha_1\cdots \alpha_d}\epsilon_{I_1\cdots I_d}\partial_{\alpha_1}\phi^{I_1}\cdots \partial_{\alpha_d}\phi^{I_d}.
\end{equation}
In Ref.~\refcite{Nicolis:2011ey} the neutral version of the theory (no $\psi$ field) is used for a time reversal breaking fluid at zero temperature. The conserved current is identified with the number density of fluid elements, and the velocity of the fluid is taken to be in the direction of the number current
\begin{equation}
u^\mu=\frac{1}{b} J^\mu, \ \ b\equiv \sqrt{-J_\mu J^\mu}.
\end{equation}
In 2+1 dimensions, the effective diffeomorphism-invariant action takes the form
\begin{equation}
S=-\int d^3x\, \sqrt{-g}\varepsilon\left(b\right). 
\end{equation}
Where $\varepsilon$ is the energy density of the fluid. To first order in derivatives the only possible term one can add to this action is parity-odd
\begin{equation}
S_{\rm odd}=\int d^3x\, f\left(b\right) \epsilon^{\mu\nu\lambda}J_\mu\partial_\nu J_\lambda.
\end{equation}
In the presence of this term, the angular momentum density is non-zero
\begin{equation}
\ell = -2 f(b) b^2.
\end{equation}
The stress tensor $T^{ij}$ obtained as a variation from the metric contains parity odd terms, but not a Hall viscosity. However, it is possible to improve the energy-momentum tensor in the non-relativistic limit if the angular momentum density is proportional to the number density, with some constant factor $\lambda_H$
\begin{equation}
\ell = \lambda_H b.
\end{equation}
This assumption implies that the `spin' density is conserved.  After the improvement, there is a Hall viscosity term with
\begin{equation}
\eta_H=\frac{1}{2}\lambda_H b=\frac{1}{2}\ell.
\end{equation}
This relation agrees with the adiabatic calculation \eqref{Hallviscspin}.

For fluids at finite temperature one can follow a similar approach, but the `number density' has to be identified with the entropy density\cite{Dubovsky:2011sj}
\begin{equation}
s=b=\sqrt{-J_\mu J^\mu}.
\end{equation}
The velocity is then aligned along the entropy current, for this reason the effective action is said to give hydrodynamics in the entropy frame. For a charged fluid one also defines the chemical potential
\begin{equation}
\mu=u^\alpha(\partial_\alpha\psi-A_\alpha).
\end{equation}
The action is a function of both the entropy density and the chemical potential
\begin{equation}
S=\int d^3x\,\sqrt{-g} G(b,\mu).
\end{equation}
The energy density and pressure are related to the thermodynamic potential $G$ through a Legendre transform
\begin{equation}
\varepsilon= \mu\frac{\partial G}{\partial \mu}-G, \ \ p= s\frac{\partial G}{\partial s}-G.
\end{equation}
The analysis of the Hall viscosity has been made for neutral\cite{Bhattacharya:2012zx} and charged\cite{Haehl:2013kra} fluids. Consistently with the derivation in the zero temperature fluid, it was found that the energy-momentum tensor obtained from the effective action does not contain a Hall viscosity term. The Hall viscosity is absent even if torsion is allowed, the spin current cancels possible contributions in the canonical energy-momentum tensor in the same way as for massive Dirac fermions.

The analysis was made under the assumption that Hall viscosity should be obtained from a local term in the action. More recently, it has been shown in Ref.~\refcite{Geracie:2014iva} that the Hall viscosity can be obtained from a Wess-Zumino-Witten term constructed with the unimodular induced metric
\begin{equation}
G^{IJ}=s^{-1} \partial_\mu\phi^I\partial^\mu\phi^J,
\end{equation}
and the volume form $\epsilon_{IJ}$. The WZW term then takes the form
\begin{equation}
S_{WZW}=f\int d^3x\, du \sqrt{-g} s\, {\rm Tr}\left(\epsilon \partial_u G G^{-1}u^\mu\partial_\mu G\right).
\end{equation}
Where $u$ is a coordinate that interpolates between the trivial metric at $u=0$ $G_{IJ}=\delta_{IJ}$ to the physical metric $G_{IJ}(\phi)$ at $u=1$. The value of the Hall viscosity is proportional to the entropy density
\begin{equation}
\eta_H=2f s,
\end{equation}
where for a neutral fluid $f$ is a fixed number and for a charged fluid it can also depend on the charge density $n$, $f=f(n/s)$.

\subsection{Hall viscosity in gauge/gravity duals}

There has been a considerable effort to apply the methods of gauge/gravity duals, aka holography, to condensed matter systems, see Refs.~\refcite{Hartnoll:2009sz,Sachdev:2010ch,Adams:2012th} for reviews of the topic.

The gauge/gravity duality is a conjecture that maps observables of a strongly coupled field theory to fields in a classical gravitational theory in higher dimensions. If the field theory  lives in $D$ dimensions and is conformal, the dual geometry is anti-de Sitter $AdS_{D+1}$ space.\footnote{Anti-de Sitter is the maximally symmetric space with a negative cosmological constant $\Lambda$.} $AdS_{D+1}$ can be foliated in flat $D$-dimensional surfaces sitting at fixed values of a radial coordinate. In a convenient choice of coordinates, the metric of $AdS_{D+1}$ has the form
\begin{equation}\label{ads}
ds^2=\frac{dr^2}{r^2}+r^2\eta_{\mu\nu}dx^\mu dx^\nu.
\end{equation}
The $x^\mu$ directions are identified with the dual field theory. The isometry
\begin{equation}
r\to \lambda r, \ x^\mu\to  x^\mu/\lambda,
\end{equation}
shows that the radial coordinate is related to changes of scale in the dual theory. Heuristically, as $r\to\infty$ the geometry describes short wavelength dynamics, while as $r\to 0$ it describes long wavelengths. The generalization to other systems without conformal invariance involves changing the geometry. In particular, systems at finite temperature are described by black holes, with the Hawking temperature of the black hole equal to the temperature of the field theory. The metric \eqref{ads} is then modified, and in particular there is an event horizon at a finite value of $r$.

Global symmetries in the field theory become local symmetries in the gravity dual. For instance the $U(1)$ particle number current in the field theory is described by a gauge field $A_M$ ($M=r,0,\cdots,d$) in the gravity dual. Spacetime symmetries become diffeomorphisms, and the energy-momentum tensor of the field theory is described by the metric $g_{MN}$ in the gravity dual. Anomalies of global currents can also be described in the gravity dual, they are simply related by descent equations. For instance, a chiral anomaly in a two-dimensional field theory will be manifested in the gravity dual as a Chern-Simons term in the action of the gauge field
\begin{equation}
S_{\rm anom}\propto \int d^3 x\,\epsilon^{MNP}A_M \partial_N A_P.
\end{equation}
In odd dimensions topological effects like the anomalous Hall conductivity map to topological terms in the bulk like the Pontryagin density
\begin{equation}
S_{\rm Hall}\propto \int d^4 x\, \epsilon^{MNPQ}F_{MN}F_{PQ}.
\end{equation}

\subsection{Holographic models with an axion}

A non-zero Hall viscosity has also been computed in a class of gauge/gravity models, starting with the original work Ref.~\refcite{Saremi:2011ab}, and extended in Refs.~\refcite{Chen2011,Chen2012,Cai:2012mg}. In these models gravity is coupled  to an axion field $\varphi$
\begin{equation}\label{hologact}
S=\frac{1}{16\pi G_4}\int d^4x\sqrt{-g}\left(R-2\Lambda-\frac{1}{2}\partial_M\varphi\partial^M\varphi-V(\varphi)+\frac{\lambda}{4}\varphi\epsilon^{MNPQ}R^A_{\ BMN} R^B_{\ APQ}\right).
\end{equation}
In this action $R$ is the Ricci scalar of the metric and $R^A_{\ BMN}$ is the Riemann tensor. $G_4$ is the Newton's constant in four dimensions. 

For an appropriate potential the solution to the equations of motion for the scalar field has a non-trivial profile depending on the radial coordinate $\varphi(r)$. In the field theory parity is broken by the expectation value of a pseudoscalar operator, although is some cases there is also explicit breaking. Physical solutions with a non-trivial scalar profile exist at finite temperature. The metric asymptotes \eqref{ads} when $r\to \infty$, but it deviates from $AdS_4$ for smaller values of $r$ and there is an event horizon. This model can be extended to describe a charged system by adding a gauge field with Maxwell's action and a possible coupling to the axion.

Correlation functions of the energy-momentum tensor can be computed  in the gravity dual from classical fluctuations of the metric and the scalar around the background solution. The equations of motion are derived from the action \eqref{hologact} and expanded in the black hole geometry to linear order. The result of this calculation is that the Hall viscosity is determined by the scalar at the horizon and the temperature
\begin{equation}
\frac{\eta_H}{s}\propto \lambda T \varphi'(r_H)\propto \left.\frac{dV}{d\varphi}\right|_{r=r_H},
\end{equation} 
where $s$ is the entropy density. The last relation implies that a massless scalar (dual to a marginal operator) cannot give rise to a Hall viscosity. There are however interesting effects at higher order in derivatives\cite{Delsate:2011qp} and there can be a spontaneous generation of angular momentum and edge currents.\cite{Liu:2012zm}

\subsection{Holographic $p$-wave model}

A different kind of model that also exhibits a non-zero Hall viscosity was recently constructed in Ref.~\refcite{Son:2013xra}. The action is Einstein gravity coupled to a $SU(2)$ gauge field.
\begin{equation}\label{hologact2}
S=\frac{1}{16\pi G_4}\int d^4x\sqrt{-g}\left(R-2\Lambda-\frac{1}{4}F_{MN}^aF^{a\,MN}\right).
\end{equation}
Where the $SU(2)$ color index takes values $a=1,2,3$. In the dual field theory there is a $SU(2)$ global symmetry with a conserved current $J_\mu^a$. When a chemical potential $\mu^3$ is introduced, the global symmetry is broken to $U(1)_3$.  Lorentz symmetry is also broken to the group of spatial rotations $SO(2)_V$ by the chemical potential and the temperature. Using the gravity model, one can show that at low enough temperatures there is a phase where the spatial components of the current acquire an expectation value
\begin{equation}
\vev{J_i^{\ a}}=\Delta\delta_i^a, \ \ i,a=1,2.
\end{equation}
This breaks spontaneously the global symmetries to a diagonal $U(1)$: $SO(2)_V\times U(1)_3\to U(1)_D$, so it can be seen as a holographic model for a $p$-wave superfluid. Note that the structure constants of $SU(2)$ form the Levi-Civita symbol $\epsilon^{abc}$. The breaking of parity necessary to have a Hall viscosity then occurs naturally through the linking of $SU(2)$ and spacetime indices, with $a=3$ identified with the time direction. On the gravity side this is achieved by a background gauge field with profile
\begin{equation}
A^a= \Phi(r)\delta^a_3\, dx^0 + A(r)\delta_i^a\, dx^i.
\end{equation}
As in the axion models, the Hall viscosity is computed using the fluctuations of the metric in the gravity dual.

An interesting property is that when a boundary in the space directions is introduced, there is an edge current carrying momentum. This can be understood as a consequence of having a non-vanishing angular momentum density $\ell$. Although there is no exact analytic result, it was observed that the value of the Hall viscosity and the angular momentum density obey approximately the same relation as in condensed matter systems, even at zero temperature
\begin{equation}
\eta_H\approx \frac{\ell}{2}.
\end{equation}

\section{Summary and outlook} \label{sec:con}

The existence of a quantized Hall viscosity (in units of the density) in Galilean invariant systems has a very solid theoretical foundation, from calculations in both microscopic and effective theories.  Read's formula for the Hall viscosity\cite{Read2009,Read2011} is valid for known examples with translation invariance
\begin{equation}
\eta_H = \frac{1}{2}\bar{s}\bar{n},
\end{equation}
where $\bar{s}$ is the orbital angular momentum per particle and $\bar{n}$ the density. In Hall fluids Kohn's theorem relates the static Hall viscosity and conductivity at non-zero momentum\cite{Hoyos2012,Bradlyn2012}
\begin{equation}
\sigma_H=\frac{\bar{n}}{m\omega_c}+\frac{q^2}{2(m\omega_c)^2}\left[\eta_H-m^2\omega_c\partial_B^2\varepsilon(B) \right]+O(q^4).
\end{equation}
Where $\omega_c=B/m$ is the cyclotron frequency, $m$ the mass of the particles and $\varepsilon$ is the energy density. One could use this relation to measure the Hall viscosity using inhomogeneous electric fields. Another proposal to measure Hall viscosity is through scattering of X-rays with phonons in topological insulators.\cite{Phonon}

The situation in relativistic systems is much less developed. At zero temperature, the only class of examples considered so far are massive Dirac fermions. Although time reversal and parity are broken, the Hall viscosity is zero. There is however another transport coefficient, the torsional Hall viscosity $\zeta_H$ found by Hughes, Leigh and Fradkin\cite{Hughes2011,Hughes2013} which is closely related
\begin{equation}
\zeta_H=\frac{m^2}{2\pi}\frac{1-{\rm sign}(m)}{2}.
\end{equation}
It is not inconceivable that more complicated models with interactions or at finite density allow for a non-zero Hall viscosity, since it is not zero at finite temperature in the axionic gauge/gravity axionic model proposed by Son and Saremi\cite{Saremi:2011ab} and also at zero temperature in the $p$-wave model\cite{Son:2013xra}.

\section*{Acknowledgements}

This work is partially supported by the Israel Science Foundation (grant 1665/10).

\end{document}